\documentclass[12pt,a4paper]{article}
\usepackage[a4paper,margin=2cm]{geometry}
\usepackage{authblk}
\usepackage{siunitx}
\usepackage{subcaption}

\usepackage{graphicx}  
\usepackage{hyperref}  
\usepackage{mhchem}    
\usepackage{xcolor}    
\usepackage{amsmath}
\usepackage{amssymb}
\DeclareMathOperator{\arcsinh}{arcsinh}

\newcommand{\beginsupplement}{%
	\setcounter{table}{0}
	\renewcommand{\thetable}{S\arabic{table}}%
	\setcounter{figure}{0}
	\renewcommand{\thefigure}{S\arabic{figure}}%
	\renewcommand{\thesection}{S\arabic{section}}%
	\renewcommand{\theequation}{S\arabic{equation}}%
	
}

\author[1]{Ivan Villani}
\author[2,3]{Samuele Fracassi}
\author[2,3]{Niccolò Traverso Ziani}
\author[2,3]{Maura Sassetti}
\author[3]{Matteo Carrega}
\author[4]{Vaidotas Mišeikis}
\author[4]{Camilla Coletti}
\author[1]{Fabio Beltram}
\author[5]{Kenji Watanabe}
\author[6]{Takashi Taniguchi}
\author[1]{Sergio Pezzini$^{*}$}
\author[1]{Stefan Heun$^{*}$}

\affil[1]{Istituto Nanoscienze–CNR, NEST-Scuola Normale Superiore, Piazza San Silvestro 12, 56127 Pisa, Italy}
\affil[2]{Dipartimento di Fisica, Università di Genova, Genova, Italy}
\affil[3]{CNR-SPIN, Via Dodecaneso 33, 16146 Genova, Italy}
\affil[4]{Center for Nanotechnology Innovation, Laboratorio NEST, Istituto Italiano di Tecnologia, 56127 Pisa, Italy}
\affil[5]{Research Center for Electronic and Optical Materials, National Institute for Materials Science, 1-1 Namiki, Tsukuba 305-0044, Japan}
\affil[6]{Research Center for Materials Nanoarchitectonics, National Institute for Materials Science, 1-1 Namiki, Tsukuba 305-0044, Japan}

\affil[*]{sergio.pezzini@cnr.it and stefan.heun@cnr.it}

\title{Scanning Gate Microscopy Modulation of Supercurrent in Graphene
	Josephson Junctions}

\begin{document}

\maketitle

\begin{abstract}
	Graphene Josephson junctions represent an excellent platform for quantum technologies, thanks to the combination of high carrier mobility, ballistic transport, and gate-tunable critical currents, preserved even under quantizing magnetic fields. Investigating the spatial distribution of supercurrent flow could be crucial for elucidating transport mechanisms and advancing the engineering of these devices. In this work, we employ a scanning gate microscope to investigate supercurrent transport in hexagonal boron nitride-encapsulated graphene Josephson junctions. We study the supercurrent modulation as a function of tip voltage bias and tip-to-sample distance and provide a complete characterization of the tip-induced modulation. Our experimental results are quantitatively consistent with numerical simulations and pave the way toward local mapping and manipulation of gate-tunable superconducting phenomena with unprecedented spatial resolution.
\end{abstract}

The proximity effect induces superconductivity in a normal weak link inserted between two superconducting leads, forming a Josephson junction (JJ).\cite{tinkham_1996} Nowadays, JJs serve as key elements for a wide range of quantum technologies and superconducting devices. \cite{Lemziakov2024, andreev, Volkov1995, Martinez2015, Strambini2014, Wiedenmann2016, Sarkar2026, Turini2022, Chieppa2025, villani2026diode_arxiv} Utilizing field-effect-tunable materials such as graphene as the weak link enables gate control of the superconducting properties.\cite{Heersche2007, Butseraen2022, Draelos2019, Telesio2022, Vischi2020, Walsch2021} JJs based on graphene (GJJs) encapsulated in hexagonal boron nitride (hBN) stand out for their large supercurrent densities, high interface transparency, and ballistic transport over micrometer-scale. \cite{Calado_2015,BenShalom2016} These features enable enticing new capabilities, such as the proximity effect in the quantum Hall regime,\cite{Amet2016, Vignaud2023, Barrier2024, Villani2025, Jang2025, Diez2026arxiv} a promising path toward the realization of hardware for topological quantum computation. \cite{Lindner2012,Stern2013,Clarke2013,Alicea2016}

Traditionally, the characterization of JJs relies on electrical transport measurements. While this approach provides valuable information on a macroscopic scale, it  prevents access to spatially resolved information.  scanning gate microscopy (SGM) has emerged as a powerful technique for spatial mapping of electrical transport in gate-tunable materials. This technique is based on a voltage-biased conductive tip of an atomic force microscope (AFM) that can be employed as a movable local gate. By measuring how the tip affects the conductance of the device, one can infer the transport properties on the local scale. Pioneering work employing SGM has been performed by Topinka \textit{et al.}\cite{Topinka2000,Topinka2001,LeRoy2002,Jura2007,Jura2009} Subsequently, SGM has been employed on semiconductors,\cite{Percebois2023, Gold2021_ee_scatt} quantum wells,\cite{Konig2013} two-dimensional electron gases (2DEGs),\cite{Iagallo2015,Gold2021,Braem2018} quantum dots,\cite{Huefner2011} and graphene.\cite{Brun2019,Gold2020,Brun_2020,Brun2021} In combination with a magnetic field, SGM enabled imaging of electron trajectories\cite{Petrovic2017}  in 2DEGs \cite{Aidala2007} and in  graphene. \cite{Bhandari2016,Bhandari_2018} The technique has also been  employed  in the quantum Hall regime. \cite{Paradiso2010,Paradiso2011,Paradiso2012backscatt,Paradiso2012incstripes,Bours2017} 

Since the critical current of GJJs is  gate-tunable,\cite{Calado_2015} SGM appears as a powerful tool for the spatial investigation of local transport properties in the superconducting regime. This potential recently inspired theoretical efforts to model the effect of an SGM tip on the induced superconductivity.\cite{Kaperek2022,Maji2024,Maji2026} However, on the experimental side,  only a handful of SGM experiments have been reported on superconducting devices.\cite{Huefner2009,Hegedus2025,Bhandari2020,Lombardi2025} Huefner \textit{et al.}~used SGM to characterize superconducting single electron transistors, \cite{Huefner2009} Hegedüs \textit{et al.}~employed SGM to study the nature of two-level system defects in qubits,\cite{Hegedus2025} and Bhandari \textit{et al.}~imaged Andreev reflections  at a single graphene-superconductor interface.\cite{Bhandari2020} Ref.~\cite{Lombardi2025} represents the first example of application of SGM  to JJs. A switching current modulation of a few nA was reported in an InSb nanoflag JJ, setting an important proof-of-concept for investigating supercurrent modulation in this class of devices.

\begin{figure*}[!htb]
	\centering
	\includegraphics[width=\linewidth]{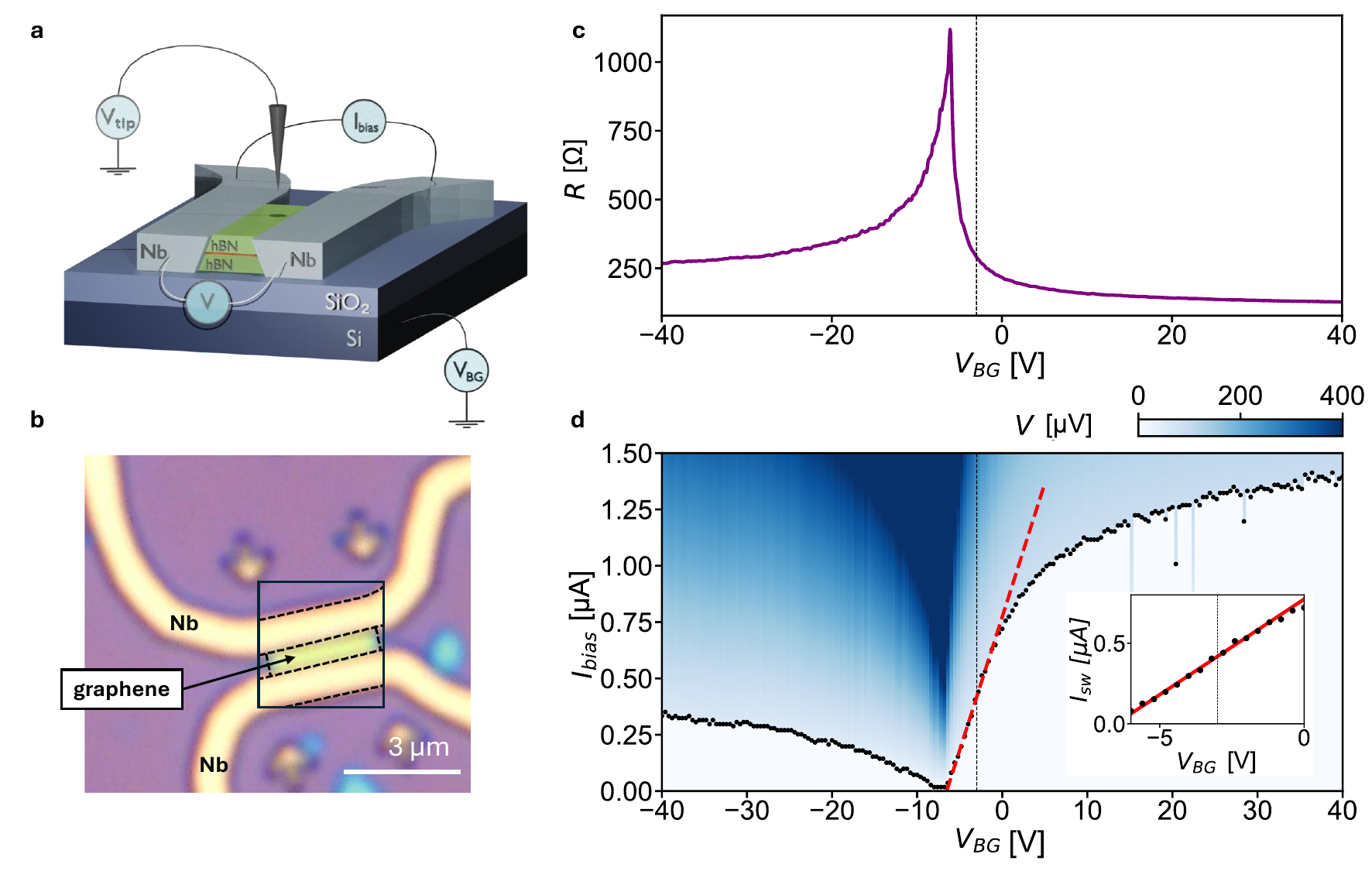}
	\caption{($\mathbf{a}$) Schematics of sample geometry and SGM setup. Graphene is encapsulated in hexagonal boron nitride (hBN) and contacted by Niobium (Nb) leads. A bias $V_{tip}$ is applied to the tip, which scans above the sample at fixed distance. A bias $V_{BG}$ is applied to the back-gate. A current $I_{bias}$ is applied to the junction, and the voltage drop $V$ is measured. ($\mathbf{b}$) Optical microscopy image of junction J1 ($W=3\;\mu$m, $L=600$~nm). The black square indicates the scanning area of SGM maps shown in Fig.~\ref{Fig2_var_Vtip}. The black dashed lines outline the junction structure. ($\mathbf{c}$) Back-gate sweep measured with a lock-in amplifier by applying a $100$~nA AC bias. The temperature is set to $T=2.7$~K to obtain the normal-state resistance $R$. The vertical dashed line indicates the  working point for the analysis presented in Fig.~\ref{Fig2_var_Vtip}, at $V_{BG}=-3$~V. ($\mathbf{d}$) Voltage drop as a function of back-gate voltage $V_{BG}$ and DC current bias $I_{bias}$. The sweep direction is from 0 to finite current values. The white region of zero voltage drop identifies the supercurrent branch. The switching current, extracted by setting a threshold of $1.5\;\mu$V on the measured voltage drop, is indicated by black dots. The dashed red line is a linear fit to $I_{sw}$ across the working point, used to determine the back-gate switching current modulation efficiency, $\beta_{BG,n}=118\pm 3$~nA/V. (\textit{Inset}) Switching current in a $\pm 3$~V range around the working point.  The dependence of $I_{sw}(V_{BG})$ can be approximated as linear. \label{Fig1_sample_overview}}
\end{figure*}

Here, we employ SGM to investigate the supercurrent modulation in hBN-encapsulated GJJs. Thanks to  large switching currents  and a geometry that is particularly suitable for SGM investigation, we obtain a large tip-induced modulation of hundreds of nA. We study the tip-induced supercurrent modulation as a function of tip voltage  (at fixed tip height) and of tip-to-sample distance. A theoretical model supporting our experimental results is also reported. We  provide a complete   characterization of the tip-induced modulation of the critical current.  This work thus identifies GJJs as the ideal platform for imaging supercurrent transport with unprecedented spatial resolution.

SGM measurements were performed  at a base temperature of $300$~mK. The sample is mounted on an Attocube AFM head attached to the \ce{^3He} pot of a Janis cryostat. The AFM-SGM tip is mounted on a tuning fork and is electrically connected to a voltage source that controls the tip voltage bias $V_{tip}$. All electrical lines, including the connections to the AFM components, are equipped with cryogenic filters. A sketch of the device layout and measurement configuration is shown in Fig.~\ref{Fig1_sample_overview}a. Graphene is encapsulated in two $30$~nm thick hBN flakes, resulting in a total stack thickness of $60$~nm. The stack is released onto a \ce{Si/SiO_2} wafer ($300$~nm oxide thickness) that acts as a back-gate. Niobium (Nb) edge-contacts are obtained by DC magnetron sputtering performed after dry etching of the hBN-graphene-hBN stack. The thickness of the Nb contacts was conveniently set to match the height of the stack,   to avoid height differences on the surface of the devices. We employ graphene single crystals grown by chemical vapor deposition \cite{Miseikis_2015} that guarantee electrical transport properties comparable to state-of-the-art exfoliation-based samples.\cite{Pezzini_2020} Further details on sample layout  are presented in the Supplementary Material. The investigated sample comprises two JJs of different length $L$ and width $W$. Junction J1 has  $L=600$~nm, $W=3\;\mu$m,  junction J2 $L=800$~nm, $W=4\;\mu$m. The two JJs are in the long to intermediate ballistic regime and exhibit comparable transport properties, both in the normal state and the superconducting regime. 

\begin{figure*}[!htb]
	\centering
	\includegraphics[width=\linewidth]{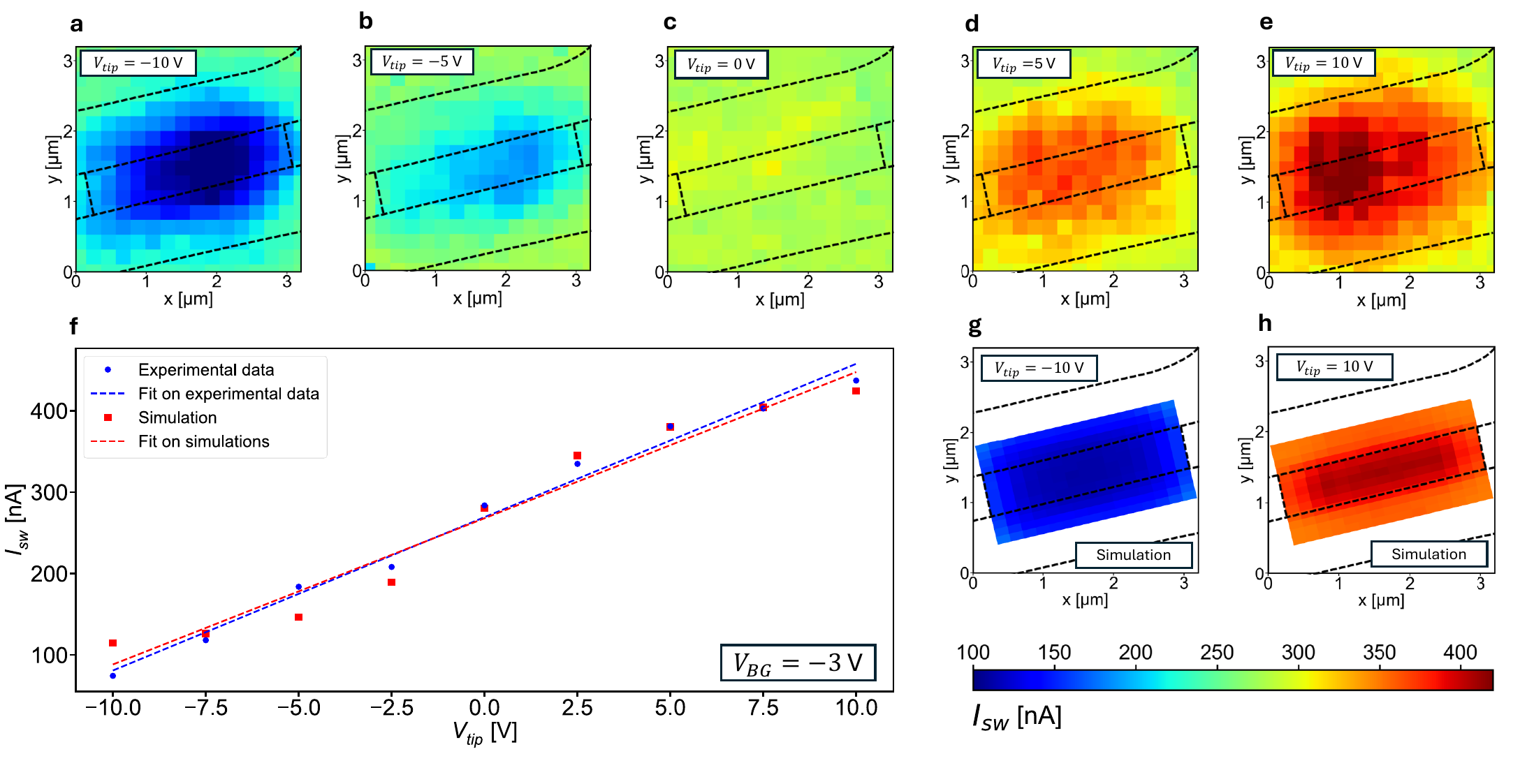}
	\caption{\label{Fig2_var_Vtip}  ($\mathbf{a}-\mathbf{e}$) SGM maps of  $I_{sw}$ in the electron doping regime ($V_{BG}=-3$~V), for different values of  $V_{tip}$. Black dashed lines outline the junction geometry. ($\mathbf{f}$) Critical current modulation as a function of $V_{tip}$. Experimental points (extracted from SGM maps as in \textbf{a}-\textbf{e}) and best fit are shown in blue: $\beta_{tip,n}=18.8\pm 0.8$~nA/V. Results of numerical simulations are shown in red and are in excellent agreement with experimental data: $\beta_{simul,n}=18.0\pm 1.4$~nA/V. ($\mathbf{g}$) Simulation of the SGM map in \textbf{a}, for $V_{tip}=-10$~V. ($\mathbf{h}$) Simulation of the SGM map in \textbf{e}, for $V_{tip}=10$~V. Same color scale  for  experimental data and simulations. Junction J1, $T=300$~mK, $d=100$~nm.}
\end{figure*}

An optical microscopy image of junction J1 is shown in Fig.~\ref{Fig1_sample_overview}b (to avoid damage to the GJJs that occurred in test experiments, no AFM topographic images were acquired during this experiment). The normal-state resistance of J1 is modulated by the back-gate voltage ($V_{BG}$) as shown in Fig.~\ref{Fig1_sample_overview}c. Maximum resistance is observed at the charge neutrality point (CNP), which is shifted towards negative $V_{BG}$ as a result of the induced doping by the Nb contacts. In junction J1, $V_{BG}^{CNP}=-6.3$~V, similar to previous devices fabricated using the same protocol.\cite{Villani2025,villani2026diode_arxiv}  The resistance approaches its minimum value to the right of the CNP, $V_{BG}>V_{BG}^{CNP}$, in the electron doping regime. For $V_{BG}<V_{BG}^{CNP}$, in the hole doping regime, the resistance is always larger due to p-n barriers originating from Fermi level pinning at the Nb-graphene interfaces.\cite{Calado_2015,BenShalom2016,Villani2025}  The induced supercurrent is also modulated by the back-gate voltage, as illustrated in Fig.~\ref{Fig1_sample_overview}d, where we report the voltage drop $V$ across J1 as a function of $V_{BG}$ and DC current bias $I_{bias}$. In these measurements, $I_{bias}$ is swept from $0$ to finite values. The supercurrent branch corresponds to the white region of zero voltage drop, and the switching current $I_{sw}$, extracted by setting a
threshold of $1.5\;\mu$V on the measured voltage drop $V$, is indicated by the black dots. $I_{sw}$ is maximum for large electron doping, lower on the hole doping side, and minimum at CNP. Overall, the switching current shows a non-linear dependence on $V_{BG}$.
A convenient choice for SGM investigation is a working point with a large $I_{sw}$, combined with a linear $I_{sw}(V_{BG})$ dependence with a large slope. The working point was set to $V_{BG}=-3$~V, in the electron doping regime, indicated in Fig.~\ref{Fig1_sample_overview}c by the black dashed line. At this working point, the number of propagating modes (for each spin and valley degree of freedom) is equal to $N=75$. The corresponding interface transparency is $Tr=0.5$, calculated following Ref.~\cite{BenShalom2016}. Performing a linear fit to $I_{sw}(V_{BG})$ across the working point (red dashed line in Fig.~\ref{Fig1_sample_overview}d), we obtain a back-gate modulation efficiency $\beta_{BG,n}=118\pm 3$~nA/V.
$I_{sw}(V_{BG})$ is well approximated by this linear dependence  (inset to Fig.~\ref{Fig1_sample_overview}d).

In Fig.~\ref{Fig2_var_Vtip} we investigate the switching current modulation versus $V_{tip}$ of junction J1. The tip is scanned in vacuum above the junction at a fixed distance of $d_{vac}=70$~nm from the top surface. Given the thickness of the top hBN flake $d_{hBN}=30$~nm, the distance between the tip and the graphene channel is $d=d_{vac}+d_{hBN}=100$~nm. In Fig.~\ref{Fig2_var_Vtip}a-e we report SGM maps of    $I_{sw}$ as a function of tip position for different $V_{tip}$ values, ranging from $-10$~V to $+10$~V. Maps are acquired by defining a grid over an area of $3.2\times 3.2\;\mu$m$^2$ (continuous black square in Fig.~\ref{Fig1_sample_overview}b). A $V$-$I_{bias}$ curve is measured at each pixel of the grid, by sweeping the DC bias from 0 to finite values, and  $I_{sw}$ is extracted with the same criterion as in Fig.~\ref{Fig1_sample_overview}d, by setting a threshold on the measured voltage drop. The obtained values are then plotted as a function of $x$-$y$ position. Black dashed lines outline the junction geometry. We observe a reduction of the switching current for negative $V_{tip}$, while we observe an enhancement for positive $V_{tip}$. This is   consistent with a positive back-gate modulation coefficient $\beta_{BG,n}>0$: in the electron doping regime, supercurrent is enhanced (reduced) as the back-gate voltage is increased (reduced). Furthermore, the switching current variation is larger as $|V_{tip}|$ is increased. For $V_{tip}=-10$~V we observe the maximum switching current reduction (Fig.~\ref{Fig2_var_Vtip}a). The reduction gets weaker for $V_{tip}=-5$~V (Fig.~\ref{Fig2_var_Vtip}b), and becomes negligible at $V_{tip}=0$~V (Fig.~\ref{Fig2_var_Vtip}c). For positive $V_{tip}$, the switching current is enhanced, and a larger increase is reported in Fig.~\ref{Fig2_var_Vtip}e ($V_{tip}=10$~V) compared to Fig.~\ref{Fig2_var_Vtip}d ($V_{tip}=5$~V). In each map, the maximum   modulation is observed when the tip is located over the junction, and  decreases when the tip is moved away.

\begin{figure*}[!t]
	\centering
	
	\includegraphics[width=\linewidth]{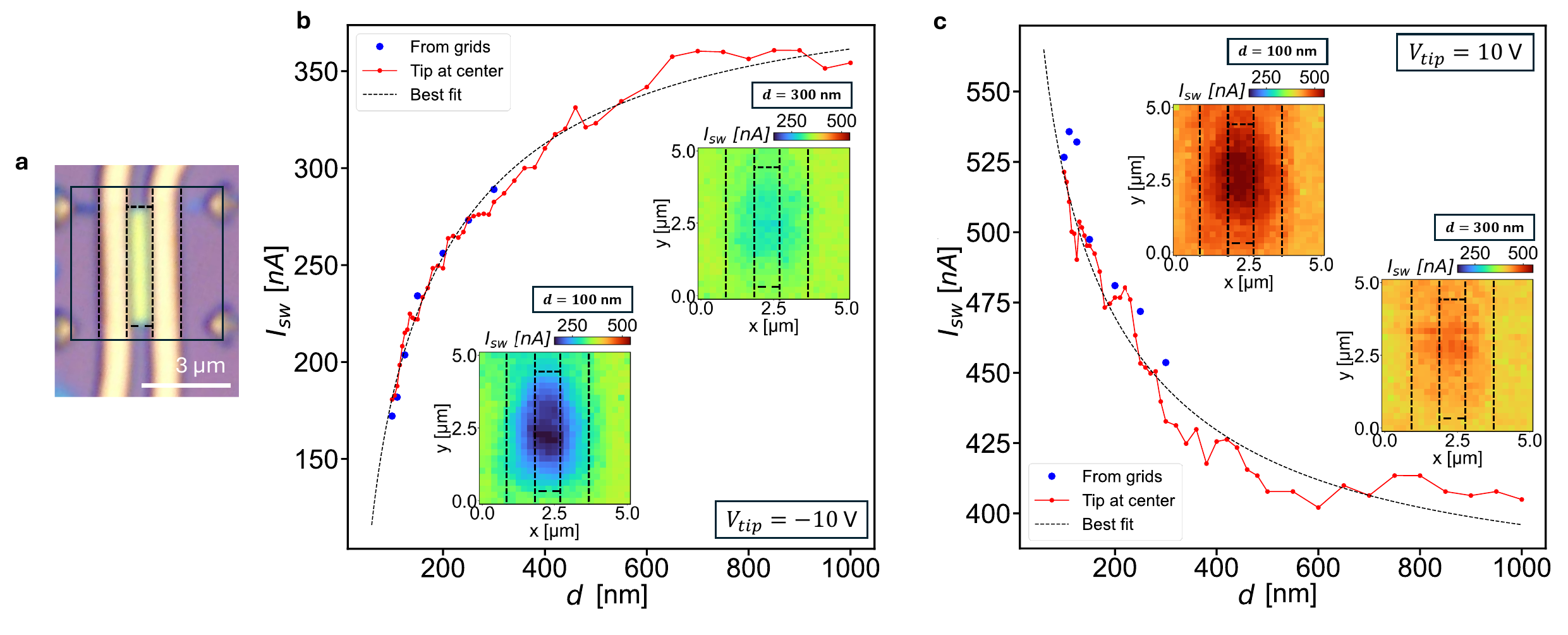}
	\caption{\label{Fig3_var_d}  ($\mathbf{a}$) Optical microscopy image of junction J2. The black square indicates the scanning area for SGM maps. Dashed black lines outline the device geometry. ($\mathbf{b}$-$\mathbf{c}$) $I_{sw}$ as a function of distance $d$ between  tip and  graphene channel in the electron doping regime ($V_{BG}=-3$~V). $V_{tip}=-10$~V in \textbf{b}, $V_{tip}=10$~V in \textbf{c}. Best fits to red dots using Eq.~\eqref{eq:fitting_Ic_of_z} are plotted as dashed black lines.  $5\times 5\;\mu$m$^2$ SGM maps of $I_{sw}$, for $d=100$~nm and $d=300$~nm, are presented in the insets. Same color scale for all four images. Junction J2, $T=300$~mK.}
\end{figure*}

In Fig.~\ref{Fig2_var_Vtip}f we show the maximum switching current modulation extracted from each SGM map as blue dots (data acquired with the tip held fixed over the junction center are reported in the Supplementary Material). For the switching current value at $V_{tip}=0$~V we consider the average value over the entire map in Fig.~\ref{Fig2_var_Vtip}c. We obtain the  tip modulation efficiency from a linear fit to $I_{sw}(V_{tip})$ (blue dashed line in Fig.~\ref{Fig2_var_Vtip}f) as  $\beta_{tip,n}=18.8\pm 0.8$~nA/V, equal to $16\pm 1\%$ of the modulation induced by the back-gate voltage $\beta_{BG,n}$ in the electron doping regime. In other terms, applying a tip bias $V_{tip}=\pm10$~V has the effect of applying a back-gate voltage of $\pm 1.6$~V relative to the working point. In this limited interval we can consider $I_{sw}$ as directly proportional to $V_{BG}$ (see inset to Fig.~\ref{Fig1_sample_overview}d). The same analysis for the hole-doping regime (at $V_{BG}=-10.6$~V) is presented in Supplementary Material. Analogous data and analysis from junction J2, reported in Supplementary Material, give consistent results.

In Fig.~\ref{Fig2_var_Vtip}f we also report results from numerical simulations of the tip-induced modulation. Simulations are performed with a coarse-grained method similar to Ref.~\cite{Lombardi2025}, that describes the tip potential at the graphene plane with an effective Lorentzian shape. \cite{Szafran2011,Herbschleb2015}   Details on the simulations are reported in Supplementary Material. Simulation results of the maximum critical current modulation  as a function of $V_{tip}$ are reported as red squares in Fig.~\ref{Fig2_var_Vtip}f (the term \textit{critical} current refers to simulated results, while \textit{switching} current to experimental results). From a linear fit (dashed red line), we obtain $\beta_{simul,n}=18.0\pm 1.4$~nA/V, in excellent agreement with the experimental value. In the experimental data points and the simulation in Fig.~\ref{Fig2_var_Vtip}f we see a marginal deviation from the linear behavior, following an "s-shape". This feature can be ascribed to the energy-dependence of the barrier at the Nb-graphene interface that is renormalized by the tip potential.\cite{Lombardi2025}

Figures~\ref{Fig2_var_Vtip}g-h show simulated SGM maps for $V_{tip}=-10$~V (Fig.~\ref{Fig2_var_Vtip}g) and $V_{tip}=10$~V (Fig.~\ref{Fig2_var_Vtip}h). The color scale is the same as for the experimental data. Both simulated SGM maps quantitatively reproduce the experimental results and capture all relevant  features.  Maximum switching current modulation is observed within the junction channel, the modulation decreases toward the junction edge, but is  not entirely suppressed when the tip is beyond the junction and located above the Nb contacts.

Interestingly, we note that in Fig.~\ref{Fig2_var_Vtip}a-e the maximum supercurrent modulation is not at the junction center. A stronger switching current reduction is observed on the right in Fig.~\ref{Fig2_var_Vtip}a, for $V_{tip}<0$~V (same   in Fig.~\ref{Fig2_var_Vtip}b), while a stronger enhancement is observed on the left part of the junction in Fig.~\ref{Fig2_var_Vtip}e. This asymmetry likely arises from a non-uniform electrostatic potential landscape in the junction, and highlights the capability of SGM to access information on spatial variations across a superconducting device.

In Fig.~\ref{Fig3_var_d} we analyze the supercurrent modulation as a function of the distance $d$  for junction J2. An optical microscopy image of  J2 is shown in Fig.~\ref{Fig3_var_d}a. The  scanning area of SGM maps is represented by the black square. We report results relative to the n-type doping regime, at $V_{BG}=-3$~V (for J2, $V_{BG}^{CNP}=-7.6$~V) for two different values of $V_{tip}$. In Fig.~\ref{Fig3_var_d}b we report    $I_{sw}$ as a function of   $d$ for $V_{tip}=-10$~V, while in Fig.~\ref{Fig3_var_d}c for $V_{tip}=10$~V. We present data obtained with two  methods: (1) we keep the tip position fixed above the junction center, and change only $d$ (red points); (2) we acquire spatial maps, such as those  in Fig.~\ref{Fig2_var_Vtip}a-e, and extract $I_{sw}$ at junction center (blue dots).  SGM  maps of the switching current at $d=100$~nm and $d=300$~nm are shown in the insets to Figs.~\ref{Fig3_var_d}b,c. The switching current modulation is maximum at minimum distance $d=100$~nm and rapidly decays when the distance is increased. We observe a reduction of $I_{sw}$ for $V_{tip}<0$~V and an enhancement for $V_{tip}>0$~V, as expected from $\beta_{BG,n}>0$.

In the following we illustrate an  analytical model, that accurately reproduces the $I_{sw}(d)$ relationship (all  details are reported in  the Supplementary Material). Following Ref.~\cite{Brun_2020}, we model the SGM tip as a biased point-like conductive tip placed in front of a conductive plane (graphene). The tip induces a change in the carrier density $\Delta n(x,y)$. Since at the working point ($V_{BG}=-3$~V) $I_{sw}$ can be considered as directly proportional to $V_{BG}$ (and to the charge carrier density $n_{BG}=C_g \left(V_{BG}-V_{BG}^{CNP}\right)$, where $C_g$ is the back-gate lever arm), it is reasonable to assume that the same linear relationship holds on the local scale, as well. This means that the local variation of critical current density $\Delta J_c(x,y)$ is proportional to the local tip-induced variation in carrier density $\Delta n_{tip}(x,y)$. The total critical current is obtained by integrating $\Delta J_c$ over the junction area. We obtain the total critical current as a function of the tip-to-sample distance $d_{vac}$:

\begin{equation}
	I_c(d_{vac})=A\arcsinh \left(\frac{L}{2(d_{vac}+\delta)}\right)+I_{c,0},
	\label{eq:fitting_Ic_of_z}
\end{equation}

\noindent where $A$ is a proportionality constant, $\delta=d_{hBN}/\epsilon_{hBN}$ takes into account the hBN relative dielectric constant $\epsilon_r=\epsilon_{hBN}/\epsilon_0>1$ ($\epsilon_0$ is the vacuum permittivity), and $I_{c,0}$ is the critical current value at large distance $d$ (where the tip has negligible effect). Equation \eqref{eq:fitting_Ic_of_z} is obtained in the limit of $W\gg L/2,d_{vac}$ and for negligible tip curvature radius. Dashed black lines in Figs.~\ref{Fig3_var_d}b,c were obtained by fitting the red data points to Eq.~\eqref{eq:fitting_Ic_of_z}. The fits show a good agreement with the experimental data in the entire range of   $d$. Numerical results of best-fits and  data from junction J1 are reported in the Supplementary Material. The analytical model  remains robust in capturing the general $I_{sw}(d)$ dependence also at large $d$  and offers a practical framework for future experiments. 

In summary, we performed SGM experiments on GJJs, and successfully showed the possibility to locally tune the switching current with the SGM tip. We observed a large switching current modulation of several hundred nA. We studied the switching current modulation as a function of the tip voltage  and performed numerical simulations that support our experimental results. In addition, we analyzed how the switching current evolves with  tip-to-sample distance, by developing an analytical model that successfully describes the experimental observations.

The switching current modulation obtained in this work is two orders of magnitude larger than in Ref.~\cite{Lombardi2025}. This was achieved by leveraging the favorable properties of GJJs for SGM investigation. GJJs support larger switching currents up to several $\mu$A. This translates not only into a larger (and ambipolar) back-gate modulation efficiency, but also into a larger tip-induced switching current tunability. In addition, the larger dielectric constant of   hBN  increases the capacitive coupling of the tip, further enhancing its effect on the supercurrent. In addition, the use of Nb edge-contacts of the same height as the hBN-graphene-hBN heterostructure allowed us to decrease the working distance from $250$~nm (for top contacts\cite{Lombardi2025}) to $100$~nm, leading to a relevant enhancement of the tip action.  Thanks to this strong tip-induced modulation, we could  precisely investigate the effect of the tip at varying distances. A further enhancement of the supercurrent modulation could be obtained by reducing the working distance. In principle, with the optimized geometry of GJJs, it should be possible to work with the tip in contact with the top hBN flake (even though this procedure is extremely challenging due to the proximity of the Nb contacts). This would also result in an increase in the achievable lateral resolution, limited by the top hBN flake thickness and the sharpness of the tip.\cite{Lombardi2025} Our analytical model thoroughly describes the experimental data. Thus, we establish graphene-based Josephson junctions as the ideal platform for spatial investigation of supercurrent transport by SGM.

We  note that SGM maps in Fig.~\ref{Fig2_var_Vtip}a-e reveal a slightly asymmetric modulation of the switching current, likely originating from a non-homogeneous electrostatic potential landscape. In Ref.~\cite{villani2026diode_arxiv} we reported supercurrent rectification in graphene Josephson junctions arising from an asymmetric scattering potential.  SGM could be employed not only to spatially investigate the shape of such potential, but also as an active tool to locally control it. 

In conclusion, this work establishes a robust foundation for advancing SGM on GJJs for precise local mapping of superconducting phenomena, even in combination with magnetic fields, that were previously inaccessible with conventional transport characterization techniques and devices.

\section*{Supplementary Material}
See the Supplementary Material for the additional details, including experimental methods, additional data of the supercurrent modulation as a function of tip bias, SGM maps of junction conductance, additional data on junction J1 in the n-type and p-type doping regimes, additional data on junction J2, details of the numerical simulations, and additional data on the supercurrent modulation as a function of tip-to-sample distance, the analytical model for supercurrent modulation as a function of tip-to-sample distance, the connection with the Lorentzian model, additional data on junction J1 and results of best-fits.

\section*{Acknowledgments}
We acknowledge Simone Traverso for useful discussions, and support from projects PRIN2022 2022-PH852L(PE3) TopoFlag and PNRR MUR Project No. PE0000023-NQSTI. K.W. and T.T. acknowledge support from the CREST (JPMJCR24A5), JST, and World Premier International Research Center Initiative (WPI), MEXT, Japan.

\section*{Data Availability Statement}
The data that support the findings of this study are available
within the article and its supplementary material.

\clearpage
\appendix
\beginsupplement
\setcounter{section}{0}
\setcounter{figure}{0}
\setcounter{table}{0}
\setcounter{equation}{0}

\begin{center}
	\section*{Supplementary Material}
\end{center}

\section{Experimental methods}

Fabrication of graphene Josephson junctions follows the protocols described in Refs.~\cite{Villani2025,villani2026diode_arxiv_for_SI}. To facilitate tip navigation during low-temperature experiments, we integrated navigation markers in the device layout, since at $T=300$~mK tip position relies only on topography maps. We patterned two arrays of alphanumeric markers and directional arrows using the same EBL mask defined for the Nb contacts. Because the markers are co-patterned with the contacts, they share the same height ($60$~nm), preventing any interference during SGM scans. The geometry of the marker array is as follows: (1) markers are patterned over a total area of $400\times 400$~$\mu$m$^2$; (2) each array has a pitch of $10\;\mu$m, and the two arrays (markers and arrows) are spatially shifted relative to one another by $5\;\mu$m in both $x$ and $y$ directions. Given that the maximum scan size at $T=300$~mK is $8\times 8\;\mu$m$^2$, at least one marker will always be present in any scan window.

Before cooldown, a first approach is performed at room temperature, in air, with the aid of a video camera. This allows to position the tip in the area of the device where markers are located. The tip is then moved toward the $(0,0)$ coordinates. After this operation, the insert is closed, evacuated, and cooled down. During cool-down, the tip can drift by a few $\mu$m, but it always remains within the marker grid.

\clearpage
\section{Additional data of the supercurrent modulation as a function of tip bias}

\subsection{SGM maps of junction conductance}

\begin{figure}[!htb]
	\centering
	\includegraphics[width=\linewidth]{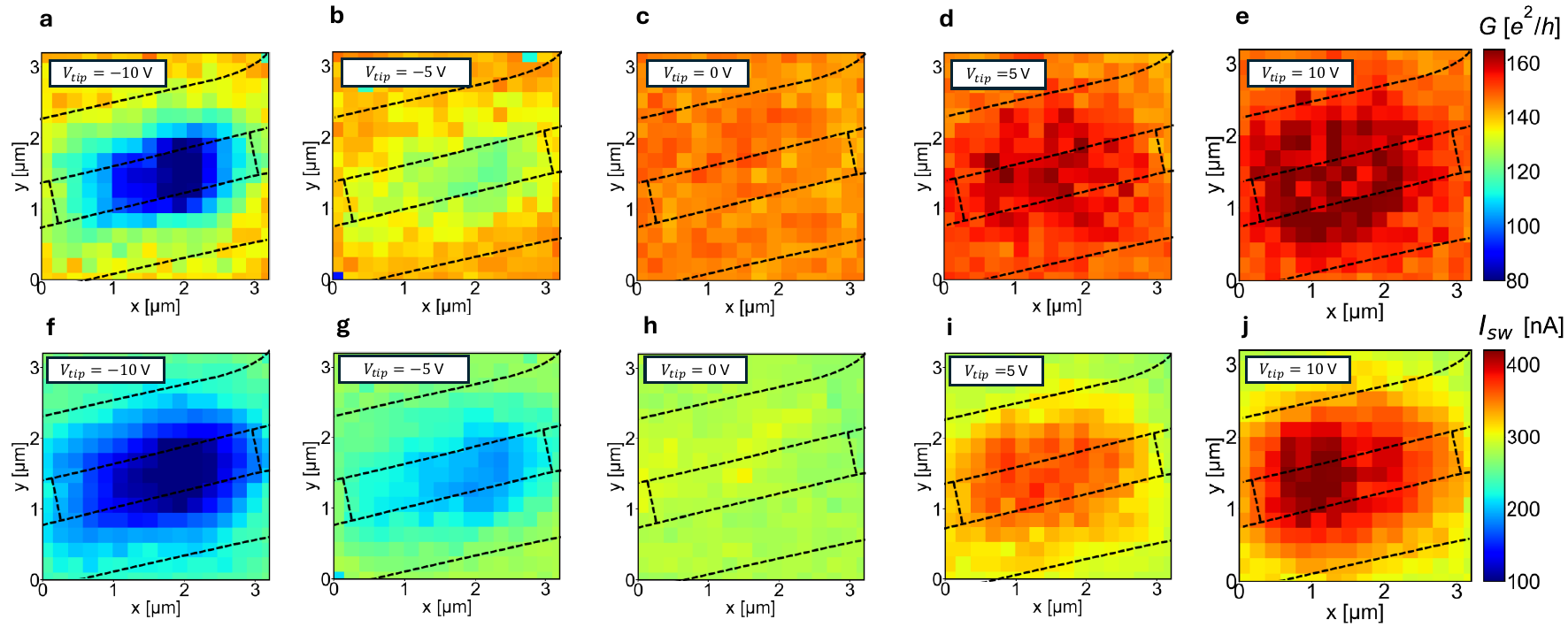}
	\caption{(\textbf{a}-\textbf{e}) SGM maps of junction conductance $G$, extracted by fitting the dissipative branch of $V$-$I$ curves. (\textbf{f}-\textbf{j}) SGM maps of the switching current $I_{sw}$, as shown in Fig.~2a-e of the main text. Same colorbar for \textbf{a}-\textbf{e}, and for \textbf{f}-\textbf{j}. \label{Gmaps_SI}}
\end{figure}

From $V$-$I$ curves used to extract the switching current shown in the SGM maps of Fig.2a-e of the main text, it is also possible to extract the sample conductance $G$ as the inverse of the slope coefficient of the fitted line in the dissipative branch. The resulting SGM maps of junction conductance are shown in Fig.~\ref{Gmaps_SI}a-e, together with the corresponding SGM of the switching current (Fig.~\ref{Gmaps_SI}f-j, same as Fig.~2a-e of the main text).

We observe a conductance modulation that directly mirrors the behavior of the switching current, characterized by a suppression of $G$ at negative $V_{tip}$ and an enhancement at positive $V_{tip}$. This corresponds to an opposite-sign variation in resistance, consistent with the negative local slope along the back-gate sweep (Fig.~1c of the main text) at the chosen working point.

\subsection{Additional data on junction J1: n-type doping regime}

\begin{figure}[!htb]
	\centering
	\includegraphics[width=\linewidth]{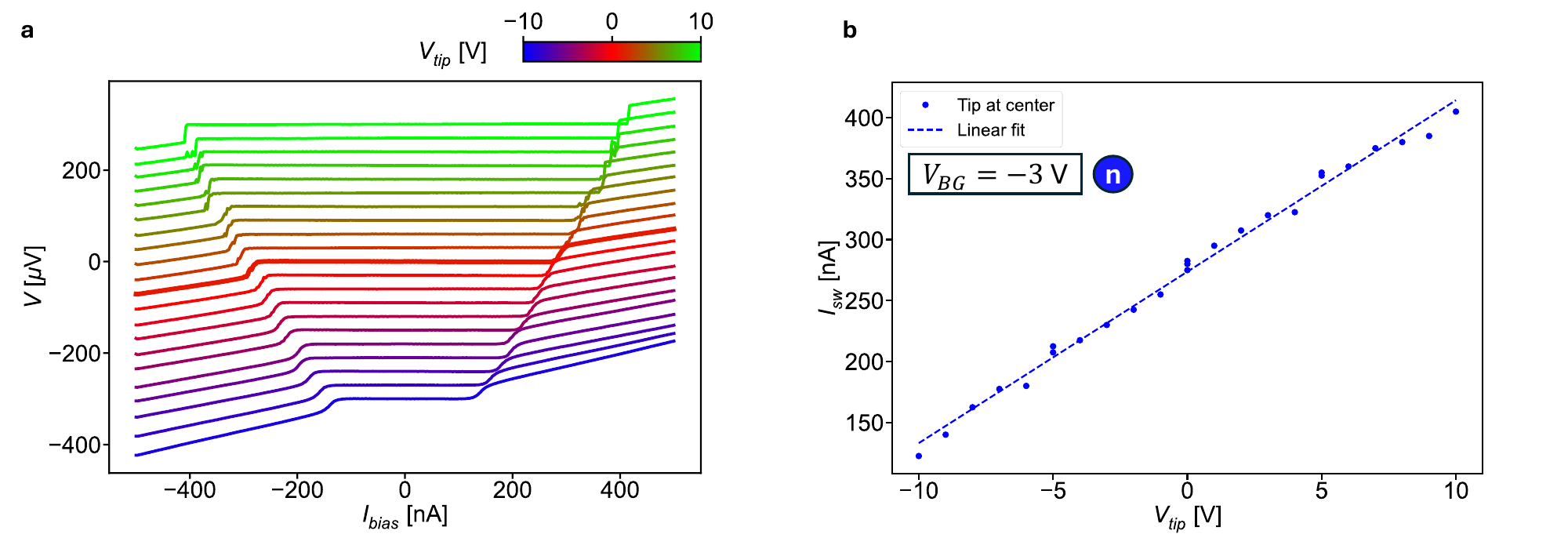}
	\caption{Supercurrent modulation in the electron doping regime ($V_{BG}=-3$~V) on junction J1, with tip positioned at junction center. ($\mathbf{a}$) Staggered $V-I_{bias}$ curves for varying $V_{tip}$. Bold line is the $V-I_{bias}$ curve at $V_{tip}=0$~V. ($\mathbf{b}$) Switching current as a function of $V_{tip}$, as extracted from panel \textbf{a}, (blue dots). The dashed blue line is a linear fit, providing a switching current modulation efficiency $\beta_{tip,n}=14.1\pm 0.3$~nA/V. \label{Fig_SI_J1_SC_var_Vtip_ndop}}
\end{figure}

In Fig.~\ref{Fig_SI_J1_SC_var_Vtip_ndop} we report additional data of switching current modulation in the electron doping regime on junction J1. A series of $V-I_{bias}$ curves (Fig.~\ref{Fig_SI_J1_SC_var_Vtip_ndop}a) was acquired by keeping the tip at the junction center at a fixed distance $d=d_{vac}+d_{hBN}=100$~nm. The extracted switching current values are plotted in Fig.~\ref{Fig_SI_J1_SC_var_Vtip_ndop}b. A linear fit gives $\beta_{tip,n}^{center}=14.1\pm 0.3$~nA/V. We note that this value is smaller than that extracted from SGM spatial maps presented in Fig.~2 of the main text ($\beta_{tip,n}=18.8\pm 0.8$~nA/V). As discussed in the main text, the critical current modulation is not symmetric around the junction center. Consequently, the maximum supercurrent modulation is not exactly at the junction center, but shifted toward one of the sample edges. Positioning the tip at the junction center thus results in a smaller critical current modulation, for both $V_{tip}>0$~V and $V_{tip}<0$~V.

\newpage
\subsection{Junction J1: p-type doping regime}

\begin{figure}[!htb]
	\centering
	\includegraphics[width=\linewidth]{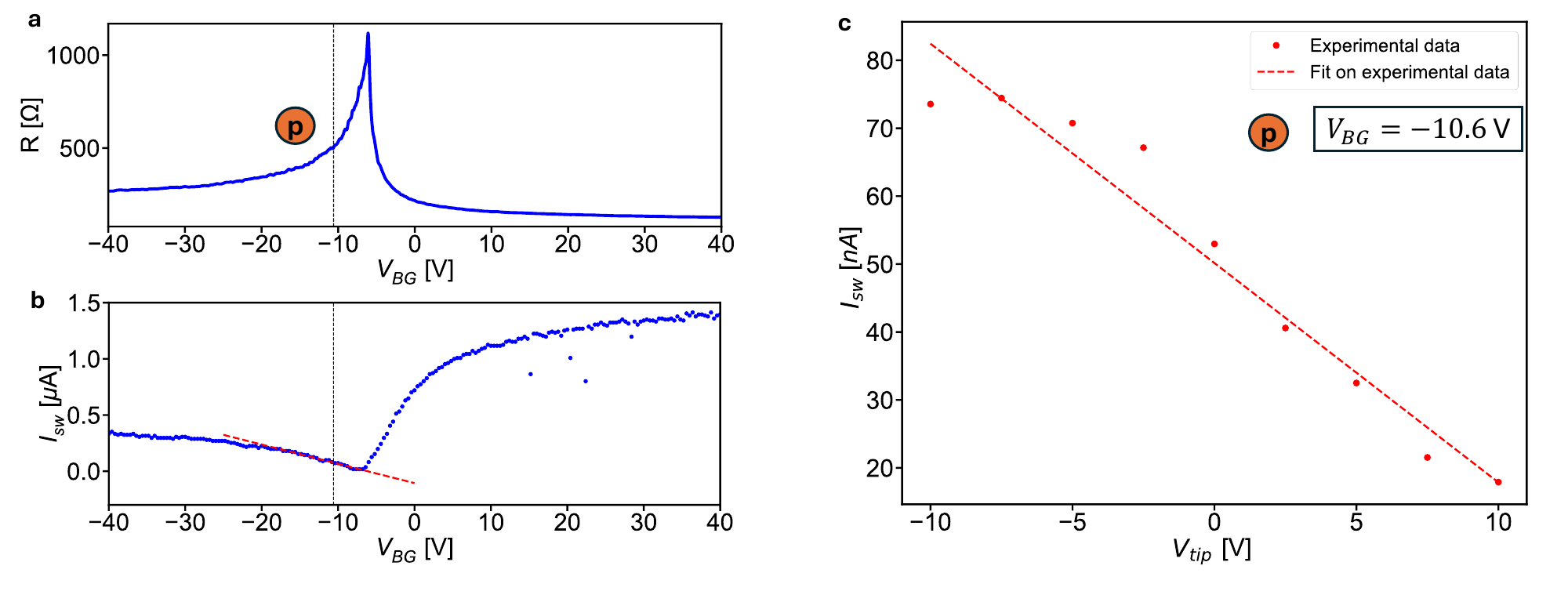}
	\caption{Supercurrent modulation in the p-type doping regime, junction J1. ($\mathbf{a}$) Back-gate sweep, with indication of the working point in the p-type doping regime (vertical black dashed line), $V_{BG}=-10.6$~V. ($\mathbf{b}$) Switching current $I_{sw}$ as a function of back-gate voltage $V_{BG}$. The dashed red line is a linear fit across the working point, giving $\beta_{BG,p}=-17\pm 2$~nA/V. ($\mathbf{c}$) Switching current as a function of $V_{tip}$ in the p-type doping regime. The calculated tip-induced modulation efficiency is $\beta_{tip,p}=-3.2\pm 0.3$~nA/V. \label{Fig_SI_J1_SC_var_Vtip_pdop}}
\end{figure}

The same analysis presented in the main text for the electron doping regime was repeated for the hole doping regime, at $V_{BG}=-10.6$~V (vertical black dashed line on the backgate sweep in Fig.~\ref{Fig_SI_J1_SC_var_Vtip_pdop}a). A linear fit passing through the working point, shown in Fig.~\ref{Fig_SI_J1_SC_var_Vtip_pdop}b, was used to extract the backgate supercurrent modulation coefficient $\beta_{BG,p}=-17\pm 2$~nA/V. The extracted critical current is plotted as a function of $V_{tip}$ (red dots) in Fig.~\ref{Fig_SI_J1_SC_var_Vtip_pdop}c, together with the resulting linear fit (dashed line) used to extract the tip induced modulation coefficient: $\beta_{tip,p}=-3.2\pm 0.3$~nA/V. Compared to the backgate modulation coefficient, $\beta_{BG,p}=-17\pm 2$~nA/V, this value is equal to $19\pm 3\%$, consistent with the percentage obtained for the electron doping case.

\subsection{Additional data on junction J2}

\begin{figure}[ht]
	\centering
	\includegraphics[width=\linewidth]{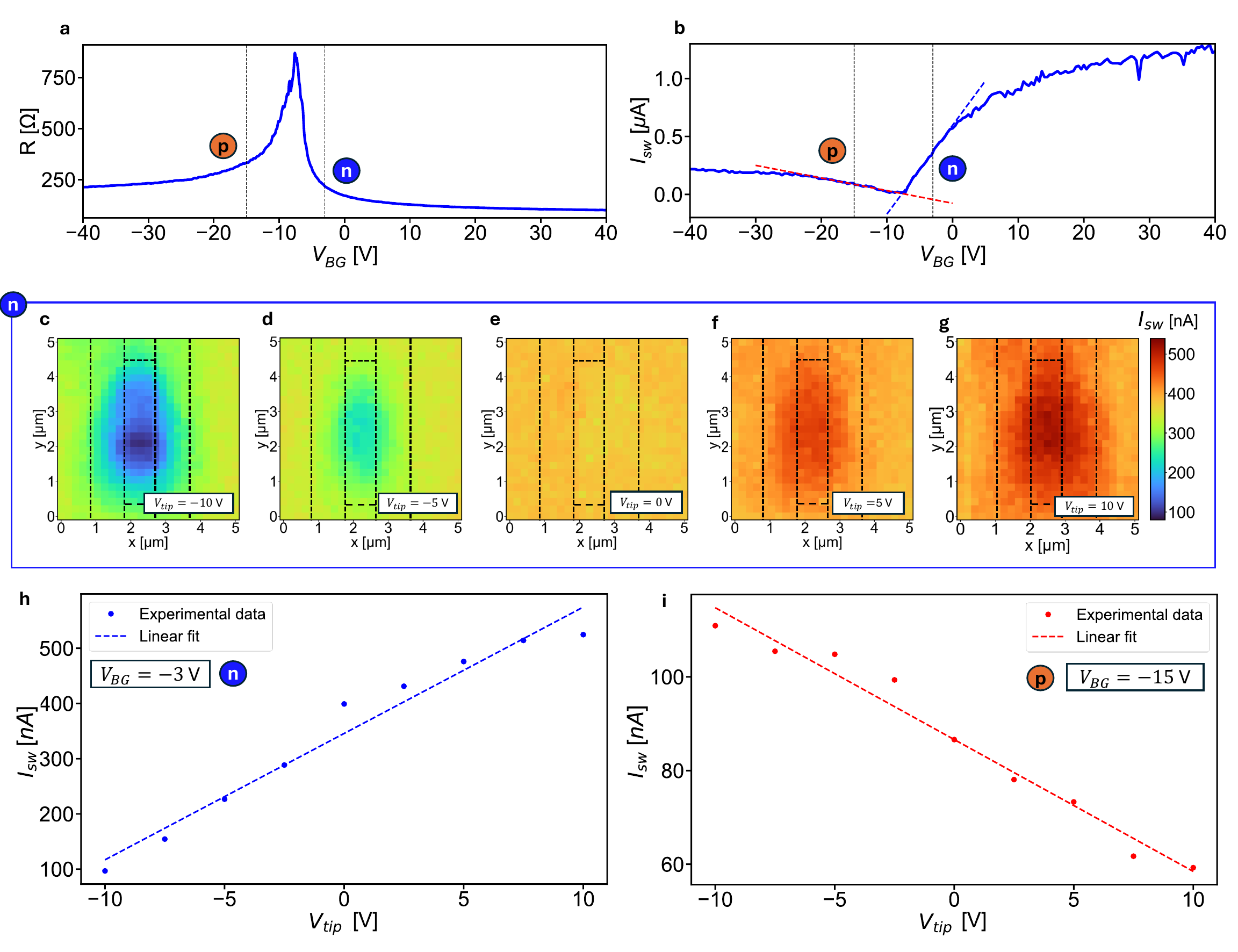}
	\caption{Supercurrent modulation as a function of tip bias $V_{tip}$, junction J2. ($\mathbf{a}$) Back-gate sweep, with indication of the working points (vertical black dashed lines): $V_{BG}=-3$~V for n-type doping, $V_{BG}=-15$~V for p-type doping. ($\mathbf{b}$) Switching current $I_{sw}$ as a function of backgate voltage $V_{BG}$. The dashed lines are linear fits across the working points, used to estimate the back-gate supercurrent modulation efficiency: $\beta_{BG,n}=76.7\pm 1.8$~nA/V (blue line), $\beta_{BG,p}=-10.9\pm 1.1$~nA/V (red line).  ($\mathbf{c}$-$\mathbf{g}$) SGM maps of the switching current in the electron doping regime ($V_{BG}=-3$~V), for different values of tip voltage bias $V_{tip}$ ranging from $-10$~V to $+10$~V. The color scale is the same for all plots. The black dashed lines outline the junction geometry. ($\mathbf{h}$-$\mathbf{i}$) Switching current as a function of $V_{tip}$. For n-type doping in $\mathbf{h}$, $\beta_{tip,n}=22.9\pm 1.7$~nA/V, for p-type doping in $\mathbf{i}$, $\beta_{tip,p}=-2.8\pm 0.2$~nA/V. \label{Fig_SI_J2_SC_var_Vtip_ndop_pdop}}
\end{figure}

Switching current modulation data as a function of tip bias $V_{tip}$ were also collected on junction J2. Results are summarized in Fig.~\ref{Fig_SI_J2_SC_var_Vtip_ndop_pdop}. Two working points were chosen similarly to junction J1, and are shown in the backgate sweep shown in Fig.~\ref{Fig_SI_J2_SC_var_Vtip_ndop_pdop}a. In the electron doping regime $V_{BG}=-3$~V, for hole doping $V_{BG}=-15$~V (for junction J2: $V_{BG}^{CNP}=-7.6$~V). The backgate supercurrent modulation efficiency is extracted from the $I_{sw}(V_{BG})$ plot of Fig.~\ref{Fig_SI_J2_SC_var_Vtip_ndop_pdop}b: $\beta_{BG,n}=76.7\pm 1.8$~nA/V, $\beta_{BG,p}=-10.9\pm 1.1$~nA/V. Representative SGM maps of the spatial modulation of the supercurrent in the electron doping regime are shown in Figs.~\ref{Fig_SI_J2_SC_var_Vtip_ndop_pdop}c-g. Maps of junction J2 were acquired by defining a 25x25 pixel grid over a $5\times 5\;\mu\text{m}^2$ area as indicated in the optical image shown in Fig.~3a of the main text. Dashed black lines in Figs.~\ref{Fig_SI_J2_SC_var_Vtip_ndop_pdop}c-g outline the junction geometry; the colormap is the same for all maps. The extracted switching current is shown in Fig.~\ref{Fig_SI_J2_SC_var_Vtip_ndop_pdop}h for electron doping, in Fig.~\ref{Fig_SI_J2_SC_var_Vtip_ndop_pdop}i for hole doping. We obtain $\beta_{tip,n}=22.9\pm 1.7$~nA/V, $\beta_{tip,p}=-2.8\pm 0.2$~nA/V, which corresponds respectively to $30\pm 2\%$ and $26\pm 3\%$ with respect to the backgate induced modulation. The absolute values of the tip-induced supercurrent modulation values are similar to junction J1, demonstrating that the tip has a comparable effect on devices with different dimensions. The back-gate supercurrent modulation efficiency values are instead lower than for junction J1, reflecting a larger degree of disorder that can be inferred from the wider resistance CNP peak in Fig.~\ref{Fig_SI_J2_SC_var_Vtip_ndop_pdop}a as compared to Fig.~\ref{Fig_SI_J1_SC_var_Vtip_pdop}a. This justifies the larger tip/back-gate modulation ratio for J2. A recap of all modulation coefficients is reported in Table \ref{tab:mod_coeff}.

\begin{table}[!htb]
	\begin{center}
		\begin{tabular}{c|cc|cc}
			& \multicolumn{2}{c|}{J1} & \multicolumn{2}{c}{J2} \\
			\hline
			Doping & $\beta_{BG}$ [nA/V] & $\beta_{tip}$ [nA/V] & $\beta_{BG}$ [nA/V] & $\beta_{tip}$ [nA/V] \\ \hline \hline
			n-doping & $+118\pm 3$ & $+18.8\pm 0.8$ & $+76.7\pm 1.8$ & $+22.9\pm 1.7$    \\
			p-doping & $-17\pm 2$ & $-3.2\pm 0.3$ & $-10.9\pm 1.1$ & $-2.8\pm 0.2$  \\
		\end{tabular}
		\caption{\label{tab:mod_coeff} Backgate- and tip-induced switching current modulation coefficients.}
	\end{center}
\end{table}

\clearpage
\newpage
\section{Details of the numerical simulations}

\begin{figure}[!h]
	\centering
	\begin{subfigure}[b]{0.48\linewidth}
		\centering
		\includegraphics[width=\linewidth]{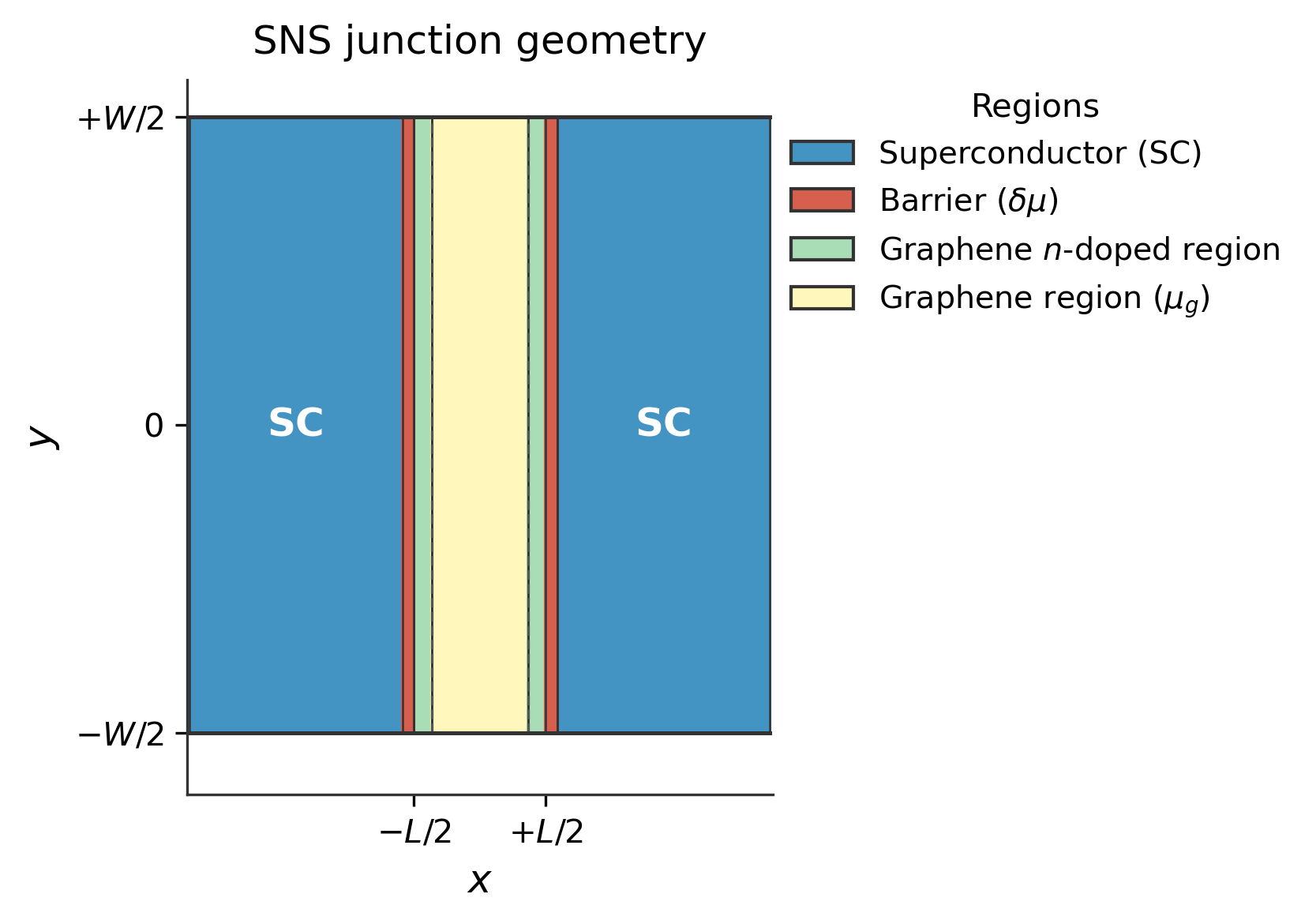}
		\caption{}
		\label{sketchSNS}
	\end{subfigure}
	\hfill
	\begin{subfigure}[b]{0.48\linewidth}
		\centering
		\includegraphics[width=\linewidth]{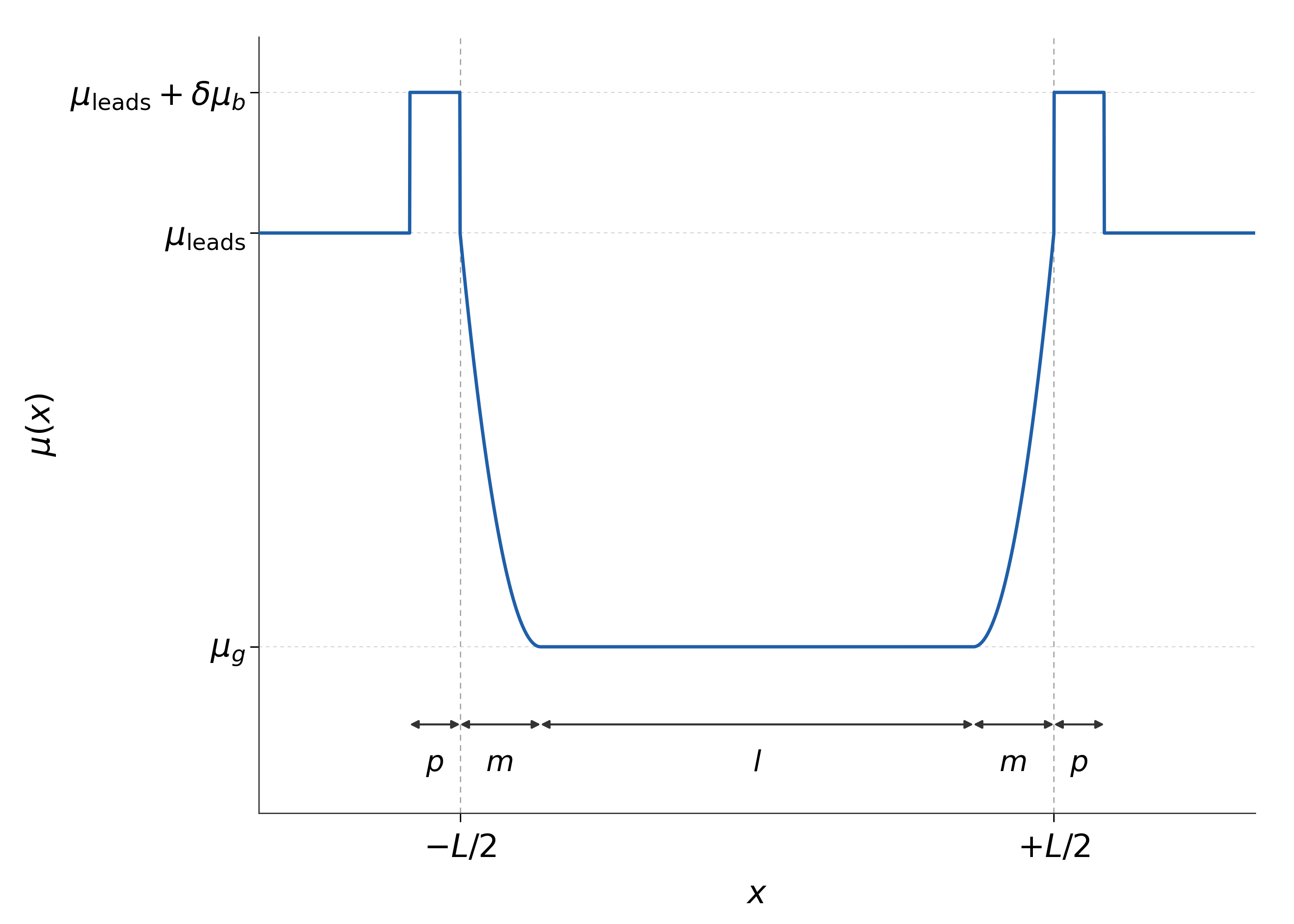}
		\caption{}
		\label{tuo_label_b}
	\end{subfigure}
	\caption{Panel \textbf{a} shows a (top view) sketch of the graphene Josephson junction. Blue regions represent the superconducting contacts with Fermi energy $\mu_{\text{leads}}$. Red areas represent the two regions at the interface with the barriers, each one of length $p\times a$. Here the chemical potential is $\mu_{\text{leads}}+\delta\mu_b$. The green regions represent the $n$-doped parts of graphene with a smoothly interpolating chemical potential $\mu(x)$. Last, the yellow part is the graphene region with flat Fermi energy $\mu_g$, tunable with the backgate voltage. Panel \textbf{b} explicitly shows the spatial dependence of $\mu(x)$ along the junction.}
	\label{fig:due_pannelli}
\end{figure}

Here we present the theoretical framework and how we performed the simulations presented in the main text. We consider a planar Josephson junction that extends in the $x-y$ plane. Defining $x$ as the longitudinal coordinate (parallel to the current flow direction), we consider a graphene lattice with armchair edges extending along $x$. Armchair edges are chosen to avoid the appearance of edge modes, which are not present in the experimental system. The graphene stripe has width $W \times a$ along $y$ and length $L \times a$ along $x$. Here $a$ is a coarse-grained length of the primitive vector. The superconducting interfaces are defined at a fixed $x$ and  extend along $y$. We define a tight-binding Bogoliubov-De Gennes (BdG) equation as reported in Ref.~\cite{Lombardi2025} and consider finite superconducting regions that extend for a length greater than the superconducting coherence length, to approximate the leads as semi-infinite. In our tight-binding model the hopping energy scale $t$ has to be renormalized because of the coarse grained $a$ \cite{Liu2014,Liu2026}. In particular, the graphene Fermi velocity of the Dirac cones $v_F \propto at$ is maintained constant. To approximate the Niobium contacts, we consider a spatial modulation of the Fermi energy, with $\mu = \mu(x)$, so that
\[ \mu(x<-L/2\times a)= \mu(x>L/2 \times a) \equiv \mu_{\text{leads}}. \]
To take the doping of the graphene by the Nb contacts into account, for $-l/2\times a < x < l/2 \times a$ we consider a region with a flat chemical potential $\mu_g$, representing the backgate-tunable Fermi energy of the graphene region, whose length is equal to $l$. At each interface with the leads, $\mu(x)$ smoothly interpolates between $\mu_{\text{leads}}$ and $\mu_g$ over a region of length $m\times a$, such that $L=2m+l$. These regions contain the $n$-doped part of graphene due to the proximity to the Niobium contacts. Moreover, to model the experimental interfaces between Niobium and graphene, we consider a barrier $\delta \mu_b$ at the interfaces that extend in the leads regions with length $p\times a$. In Fig. \ref{sketchSNS} we show a sketch of the junction with the various chemical potential zones, as discussed above (panel \textbf{a}), together with a plot of $\mu(x)$ (panel \textbf{b}).
\\
The last term that influences $\mu(x,y)$ is disorder that we model as 
\[ \mu_{dis}(x,y) = \sum_{n=1}^N v_{n}\exp\left(-\frac{(x-x_n)^2+(y-y_n)^2}{\xi^2}\right), \]
which is a sum of Gaussian puddles. Parameter $v_n$ is randomly distributed with a Gaussian distribution probability, whose standard deviation is $V_{dis}$. Each puddle has a correlation length $\xi$ and is centered at $(x_n,y_n)$. The centers are randomly extracted in the graphene region, and their number is defined such that $N\pi \xi^2/WL \equiv f $, i.e., the fraction of the area covered by disorder on the graphene sheet is fixed to $f$.
To model the presence of the SGM tip, we assume that the tip induces a local modification of the Fermi energy background. In particular, it is well known \cite{Szafran2011,Herbschleb2015,Lombardi2025} that the tip potential is captured by an effective Lorentzian shape of the form:

\begin{equation}
	\mu_{tip}(x,y) = \frac{\mu_{0,tip}}{1+\frac{x^2+y^2}{R_{Lor}^2}},
	\label{eq_SI:Lorentzian}
\end{equation}
\noindent where $R_{Lor}$ is the effective tip radius corresponding to the HWHM of the Lorentzian. Note that this parameter is different from the real tip curvature radius.
Since the tip potential renormalizes the barrier strength, which in the experiment is nearly independent of $V_{tip}$ due to screening by the Nb, we introduced a $V_{tip}$-dependent $\delta \mu_b=\delta \mu_b(V_{tip})$ to compensate the influence of $\mu_{tip}$ on the barrier strength. Specifically, we adopted a linear dependence
\[ \delta \mu_b(V_{tip}) = \delta \mu_{b,0} - \gamma V_{tip} \]
In order to estimate the Fermi energy of the graphene region $\mu_g$ as a function of experimental backgate voltage, we consider that the induced charge density is a linear function of $V_{bg}$. For graphene the charge number per unit area is $n = \frac{\mu^2}{\pi(\hbar v_F)^2}$. This leads to $\frac{d\mu_g^2(V_{bg})}{dV_{bg}} = \text{constant},$
and so 
\[ \mu_g(V_{bg}) = \alpha\,\text{sgn}(V_{bg})\,\sqrt{|V_{bg}|}. \]
Following the same argument, we have
\[ \mu_{0,tip}(V_{tip}) = \beta\,\text{sgn}(V_{tip})\,\sqrt{|V_{tip}|}.  \]
These connections completely define how to model the experimental setup. Last, to compute the supercurrent of a Josephson junction we use a thermodynamical relation, namely
\[ I_s(\phi) = \frac{2e}{\hbar}\frac{\partial E_G}{\partial \phi}, \]
where $\phi$ is the superconducting phase difference and $E_G = \sum_{E<0}E$ is the ground state energy in the BdG formalism. Once extracted the current phase relation $I_s(\phi)$, the critical current is calculated as $I_c = \text{Max}_{\phi}I_s(\phi)$. We perform the simulations using KWANT \cite{Groth_2014}. The values of the parameters are listed in Table \ref{tab:params}.
\begin{table}[ht]
	\centering
	\caption{Summary of simulation parameters.}
	\begin{tabular}{llll}
		\hline
		\hline
		Parameter & Symbol & Value\\
		\hline
		Primitive vector length & $a$ & \SI{14}{\nano\meter} \\
		Hopping energy & $t$ & \SI{67.5}{\milli\electronvolt} \\
		Sample width & $W$ & \SI{3}{\micro\meter}\\
		Sample length & $L$ & \SI{600}{\nano\meter} \\
		Normal region length & $l$ & \SI{440}{\nano\meter}\\
		Interface length & $m$ & \SI{80}{\nano\meter}\\
		Barrier length & $p$ & \SI{50}{\nano\meter}\\
		Leads Fermi energy & $\mu_{\text{leads}}$ & \SI{64}{\milli\electronvolt} \\
		Conversion $\mu_g-V_{bg}$ & $\alpha$ & \SI{27}{\milli\electronvolt\times\volt^{-1/2}}\\
		Conversion $\mu_{0,tip}-V_{tip}$ & $\beta$ & \SI{21.6}{\milli\electronvolt\times\volt^{-1/2}}\\
		Barrier strength & $\delta\mu_{b,0}$ & \SI{15}{\milli\electronvolt} \\
		Barrier renormalization coefficient & $\gamma$ & \SI{0.945}{\milli\electronvolt\times\volt^{-1}}\\
		Tip effective radius & $R_{Lor}$ & \SI{436}{\nano\meter}\\
		Standard deviation of disorder strength & $V_{dis}$ & \SI{1.35}{\milli\electronvolt}\\
		Disorder correlation length & $\xi$ & \SI{60}{\nano\meter} \\
		Disorder filling fraction & $f$ & 10\% \\
		Superconducting gap & $\Delta$ & \SI{1}{\milli\electronvolt} \\
		\hline
		\hline
	\end{tabular}
	\label{tab:params}
\end{table}

\clearpage
\newpage
\section{Supercurrent modulation as a function of tip-to-sam\-ple distance}

\subsection{Analytical model for supercurrent modulation as a function of tip-to-sample distance}

\begin{figure}[!htb]
	\centering
	\includegraphics[width=0.5\linewidth]{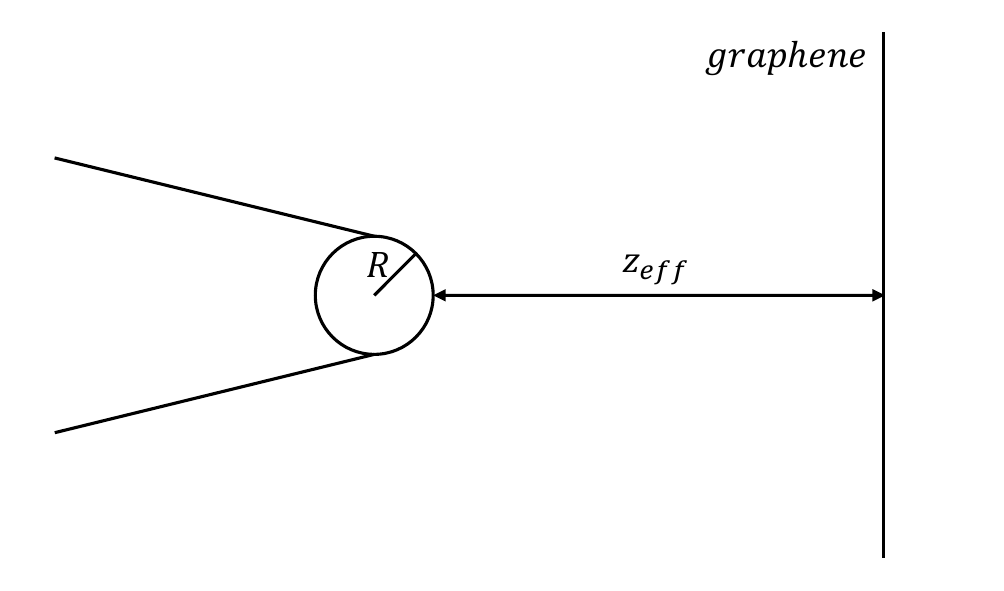}
	\caption{Schematics for the analytical model: tip in front of the graphene plane. \label{Fig_model}}
\end{figure}

Here we discuss the details of the analytical model presented in  the main text that describes the variation of the switching current as a function of the tip-to-sample distance. We start by modeling the tip as a conductive sphere with curvature radius $R$, located at distance $d=R+d_{vac}+d_{hBN}$ from the graphene sheet, as shown in Fig.~\ref{Fig_model}. $d_{hBN}$ indicates the thickness of the top hBN flake (which is fixed), $d_{vac}$ is the "thickness" of the vacuum, i.e., the distance between the tip and the top hBN flake. $d_{vac}$ changes as the tip height is varied, while $d_{hBN}$ is constant. 

\noindent We model the vacuum-hBN stack as an only-vacuum medium with effective thickness $z_{eff}$:
\[   z_{eff}=d_{vac}+\frac{d_{hBN}}{\epsilon_{hBN}}=d_{vac}+\delta, \]
where $\epsilon_{hBN}$ is the dielectric constant of hBN. Now, our system behaves as a vacuum layer with $\epsilon_0$ of thickness $z_{eff}$. 

Following Ref.~\cite{Brun_2020}, for a point-like tip (i.e. with a negligible radius $R$), due to the Dirac nature of electrons, the tip-induced charge carrier density variation is related to the potential as follows:

\[
\Delta n(z_{eff},r)=\frac{V(z_{eff},r)^2}{(\pi \hbar v_F)^2}=\left(\frac{V_0}{\pi \hbar v_F}\right)^2\frac{1}{r^2+z_{eff}^2},
\]

\noindent where the electrostatic potential $V(z_{eff},r)$ is:

\begin{equation}
	V(z_{eff},r)=\frac{V_0}{\sqrt{r^2+z_{eff}^2}}.
	\label{eq_SI:analyt_pot}
\end{equation}

Around the selected working point the total critical current can be approximated as a linear function of the backgate voltage, which is also linearly proportional to the charge carrier density. It is reasonable to assume that the linear relationship between critical current and carrier density still holds locally, so that the local variation of the critical current density $\Delta J_c(x,y)$ is proportional to $n$: $\Delta J_c(x,y)\propto \Delta n(x,y)$. The total tip-induced critical current variation is obtained by integrating over the spatial extension of the junction:

\begin{eqnarray*}
	\Delta I_c(z_{eff}) & \propto & \int_{-L/2}^{L/2}\text{d}x \int_{-W/2}^{W/2}\text{d}y \; n(x,y)=C \int_{-L/2}^{L/2}\text{d}x \int_{-W/2}^{W/2}\text{d}y \frac{1}{x^2+y^2+z_{eff}^2} \\
	& \simeq & C\frac{\pi}{2}\int_{-L/2}^{L/2}\text{d}x\frac{2}{\sqrt{x^2+z_{eff}^2}}\propto\arcsinh\left(\frac{L}{2z_{eff}}\right),
\end{eqnarray*}
where we used the following approximation for the known integral:

\[
\int_{-W/2}^{W/2} \frac{\text{d}y}{x^2+y^2+z_{eff}^2}=\frac{2}{\sqrt{x^2+z_{eff}^2}}\arctan\left(\frac{W}{2\sqrt{x^2+z_{eff}^2}}\right)\simeq \frac{\pi}{2}\frac{2}{\sqrt{x^2+z_{eff}^2}}.
\]
Here we consider $W\gg x,z_{eff}$ ($\max(x)=L/2$, $\max(z_{eff})=\max(d)$ for J2) and approximate $\arctan(...)\simeq \pi/2$. In the end, the tip-induced variation of the critical current is:

\begin{equation}
	I_c(d_{vac})=A\arcsinh \left(\frac{L}{2(d_{vac}+\delta)}\right)+I_{c,0},
	\label{eq_SI:fitting_Ic_of_z}
\end{equation}

\noindent where $A$ is a proportionality constant and $I_{c,0}$ is the critical current value at large distance $d$ (where the tip has negligible effect regardless of the applied $V_{tip}$).

\subsection*{Connection with Lorentzian model}
For the numerical simulations presented in the main text we used a Lorentzian profile for the tip-induced local variation of the Fermi energy, as illustrated in Section S3. The Lorentzian of Eq.~\eqref{eq_SI:Lorentzian} does not provide an explicit analytical dependence on the distance, but it is possible to link the two approaches -- so to find the dependence of the effective Lorentzian parameters on $z_{eff}$ -- by performing a perturbative expansion on $r^2/R_{Lor}^2$ for the Lorentzian, and on $r^2/z_{eff}^2$ for the phenomenological model presented above. The expansion of the Lorentzian of Eq.~\eqref{eq_SI:Lorentzian} gives:
\[ \mu_{tip}(r) = \frac{\mu_{tip,0}}{1+r^2/R_{Lor}^2} \simeq \mu_{tip,0}\left(1-\frac{r^2}{R_{Lor}^2}\right). \]
The phenomenological model gives (see Eq.~\eqref{eq_SI:analyt_pot}):
\[ V(z_{eff},r) \propto \frac{1}{\sqrt{r^2+z_{eff}^2}} \simeq \frac{1}{z_{eff}}\left(1-\frac{1}{2}\frac{r^2}{z_{eff^2}}\right), \]

We can then extract the dependence of $\mu_{tip,0}, R$ of Eq.~\eqref{eq_SI:Lorentzian} on $z_{eff}$, namely:
\[ \mu_{tip,0} \propto 1/z_{eff}, \]
\[ R_{Lor} \propto z_{eff}. \]

\subsection{Additional data on junction J1 and results of best-fits}

\begin{figure}[!htb]
	\centering
	\includegraphics[width=\linewidth]{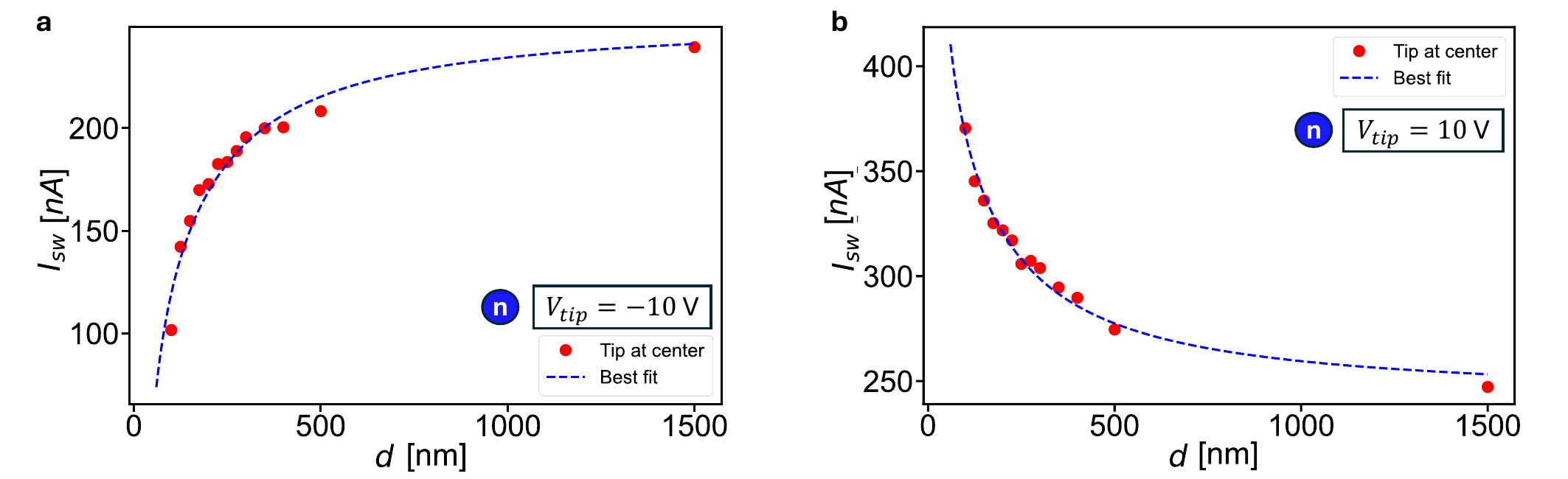}
	\caption{Switching current modulation as a function of the tip-to-sample distance $d$ in the electron doping regime, $V_{BG}=-3$~V. Data are from junction J1. ($\mathbf{a}$) $V_{tip}=-10$~V. ($\mathbf{b}$) $V_{tip}=10$~V. \label{Fig_SI_J1_SC_var_d}}
\end{figure}

In Fig.~\ref{Fig_SI_J1_SC_var_d} we report additional data of the switching current as a function of the distance $d$ between tip and graphene channel from junction J1, in the electron doping regime. Best-fits were performed using Eq.~\eqref{eq_SI:fitting_Ic_of_z}, analogously to data shown in the main text. For each junction, comparable values of the curvature coefficient $A$ and the switching current at large distance $I_{c,0}$ are obtained for $V_{tip}=\pm 10$~V (see Eq.~\eqref{eq_SI:fitting_Ic_of_z}), as summarized in Table \ref{tab:table_C}.

\begin{table}[hb]
	\begin{center}
		\begin{tabular}{c|cc|cc}
			& \multicolumn{2}{c|}{$A$ [nA]} & \multicolumn{2}{c}{$I_{c,0}$ [nA]} \\ \hline
			$V_{tip}$ & J1 & J2 & J1 & J2 \\ \hline \hline
			$-10$~V & $-66\pm 11$ & $-94\pm 5$ & $254\pm 10$ & $399\pm 6$ \\
			$+10$~V & $62\pm 7$  & $64\pm 6$  & $241 \pm 6$ & $370\pm 8$ \\
			
		\end{tabular}
		\caption{\label{tab:table_C} Best-fit parameters $A$ and $I_{c,0}$ of Eq.~\eqref{eq_SI:fitting_Ic_of_z} for junctions J1 and J2.}
	\end{center}
\end{table}

\clearpage
\bibliographystyle{unsrt}
\bibliography{Bibliography_SGM_SCmod1}

\end{document}